\documentclass[aps, superscriptaddress, floatfix, reprint]{revtex4-1}
\usepackage{graphicx}
\usepackage{dcolumn}
\usepackage{bm}
\usepackage{amsfonts}
\usepackage{amsmath}
\usepackage{amssymb}
\usepackage{color}
\usepackage[normalem]{ulem}
\usepackage[percent]{overpic}

\usepackage[colorlinks=true,citecolor=blue]{hyperref}
\hypersetup{colorlinks=true,citecolor=blue,linkcolor=red,urlcolor=blue}

\def\be{\begin{equation}}
\def\ee{\end{equation}}

\def\rv{{\bf r}}
\def\kv{{\bf k}}

\begin{document}

\title{Optical conductivity of three and two dimensional topological nodal line semimetals}
\author{Shahin Barati}
\affiliation{Department of Physics, Institute for Advanced Studies in Basic Sciences (IASBS), Zanjan 45137-66731, Iran}
\author{Saeed H. Abedinpour}
\email{abedinpour@iasbs.ac.ir}
\affiliation{Department of Physics, Institute for Advanced Studies in Basic Sciences (IASBS), Zanjan 45137-66731, Iran}
\affiliation{School of Nano Science, Institute for Research in Fundamental Sciences (IPM), Tehran 19395-5531, Iran}
\date{\today}

\begin{abstract}
The peculiar shape of the Fermi surface of topological nodal line semimetals at low carrier concentrations results in their unusual optical and transport properties.   
We analytically investigate the linear optical responses of three and two-dimensional nodal line semimetals using the Kubo formula. 
The optical conductivity of a three-dimensional nodal line semimetal is anisotropic. Along the axial direction (\textit{i.e.}, the direction perpendicular to the nodal-ring plane), the Drude weight has a linear dependence on the chemical potential at both low and high carrier dopings.  For the radial direction (\textit{i.e.}, the direction parallel to the nodal-ring plane), this dependence changes from linear into quadratic in the transition from low into high carrier concentration. 
The interband contribution into optical conductivity is also anisotropic. In particular, at large frequencies, it saturates to a constant value for the axial direction and linearly increases with frequency along the radial direction. 
In two-dimensional nodal line semimetals, no interband optical transition could be induced and the only contribution to the optical conductivity arises from the intraband excitations. The corresponding Drude weight is independent of the carrier density at low carrier concentrations and linearly increases with chemical potential at high carrier doping.
\end{abstract}
\maketitle

\section{Introduction}\label{introduction}
A remarkable progress in contemporary condensed matter physics has been the discovery of topological phases of matter and the last few years have witnessed tremendous interest in investigating these topological materials. 
Topological insulators (TIs), with their gapped bulk, and symmetry protected gapless surface states have been extensively explored, theoretically and experimentally~\cite{Hasan, Zhang,hasan_annrev2011,bernevig_book}. Recently, the main focus has been moving towards topological semimetals (TSMs), which are characterized by their protected gapless bulk states~\cite{wan_prb2011, burkov_prb2011, burkov_natmat2016}. 
In topological Weyl~\cite{burkov_prl2011, xu_prb2016} and Dirac~\cite{Young, Zhijun} semimetals, the conduction and valence bands touch each other in isolated points (\textit{i.e.}, nodal points) in the Brillouin zone (BS), and electrons have linear dispersions around these nodal points. Three-dimensional (3D) Dirac semimetals have four-fold degenerate point nodes. When the inversion (P) or time-reversal (T) symmetry is broken, each Dirac node splits into two Weyl nodes,  and the system becomes Weyl semimetal with two-fold degenerate
point nodes~\cite{Guolin}.

Much younger members of the family of TSMs are nodal line semimetals (NLSMs), specified by their one-dimensional band touchings~\cite{fang_chinphys2016,burkov_prb2011,fang_prb2015}. In NLSMs the cross-section of conduction and valence bands form either open lines, or closed loops (\textit{i.e.}, nodal rings) in the BZ~\cite{fang_chinphys2016}. This band touching is essentially non-accidental and is protected by some symmetries of the system. Among several theoretical proposals are the combined P and T or P and particle-hole (C) symmetries~\cite{zhao_prl2016}, but a full topological classification of these materials still needs to be done.
Breaking one of these underlying symmetries either fully gapes the nodal line or gapes it into several nodal points. 
It has been also proposed that tunable Floquet Weyl points could be induced in 3D NLSMs through illuminating them with circularly polarized lights~\cite{narayan_prb2016, yan_prl2016}. 

Observation of NLSMs have been theoretically envisioned in many different systems. Superlattices made of TIs~\cite{burkov_prb2011,phillips_prb2014}, hyperhoneycomb lattices~\cite{mullen_prl2015}, body-centered orthorhombic C$_{16}$ structure~\cite{wang_prl2016}, CaP$_3$~\cite{xu_prb2017}, and Ca$_3$P$_2$~\cite{chan_prb2016} compounds are just a few examples of these theoretical proposals.
Very recently, search for nodal lines in two dimensional (2D) and quasi-2D systems have been raising~\cite{li_prb2016,lu_chinphyslett,feng_arxiv2016,niu_arxiv2017_1,niu_arxiv2017_2}.   

The NLSMs, in contrast to most other topological phases,  generally lack protected surface states. This makes their experimental identification quite challenging.  As the shape of the Fermi surface of NLSMs at low chemical potentials is very peculiar and strongly anisotropic, one would naturally expect some characteristic futures of this in their transport and optical properties.  
Motivated by this, in the present paper we have employed the Kubo formula to investigate the linear optical response of nodal line semimetals. We were able to obtain analytic expressions for the intraband (\textit{i.e.}, Drude) and interband contributions into the optical conductivities of three and two-dimensional NLSMs at finite chemical potential. 
The optical properties of Dirac and Weyl semimetals have been theoretically investigated~\cite{ashby_prb2014, tabert_prb2016_1, tabert_prb2016_2, chan_prl2016}. 
The optical response of 3D nodal line semimetals has been also addressed very recently, numerically~\cite{ahn_arxiv2017}, or by considering isotropic model Hamiltonians~\cite{carbotte_jpcm2017}. 
Our main focus here instead have been on the anisotropy of the electronic structures, and we have analytically investigated the effects of frequency, chemical potential, as well as finite temperature and impurity scattering on different contributions to the optical conductivity of both three and two-dimensional NLSMs.

The rest of this paper is organized as follows. In section~\ref{sec:model} we introduce our effective low energy model Hamiltonian for a three-dimensional nodal line semimetal and discuss its different properties. In section~\ref{sec:kubo} we describe the Kubo formula for linear optical conductivity, and use it to obtain the intraband and interband contributions to the optical conductivity of a 3D NLSM along its radial and axial directions. We discuss a model 2D NLSM and its optical conductivity in section~\ref{sec:2D-NLSM}. Finally, we summarize and conclude in section~\ref{sec:summary}. 
The carrier density dependence of the Fermi energy is presented in an Appendix~\ref{sec:fermi}.

\section{Model Hamiltonian of 3D nodal line semimetals}\label{sec:model}
\begin{figure}
\centering
   \includegraphics[width=0.9\linewidth]{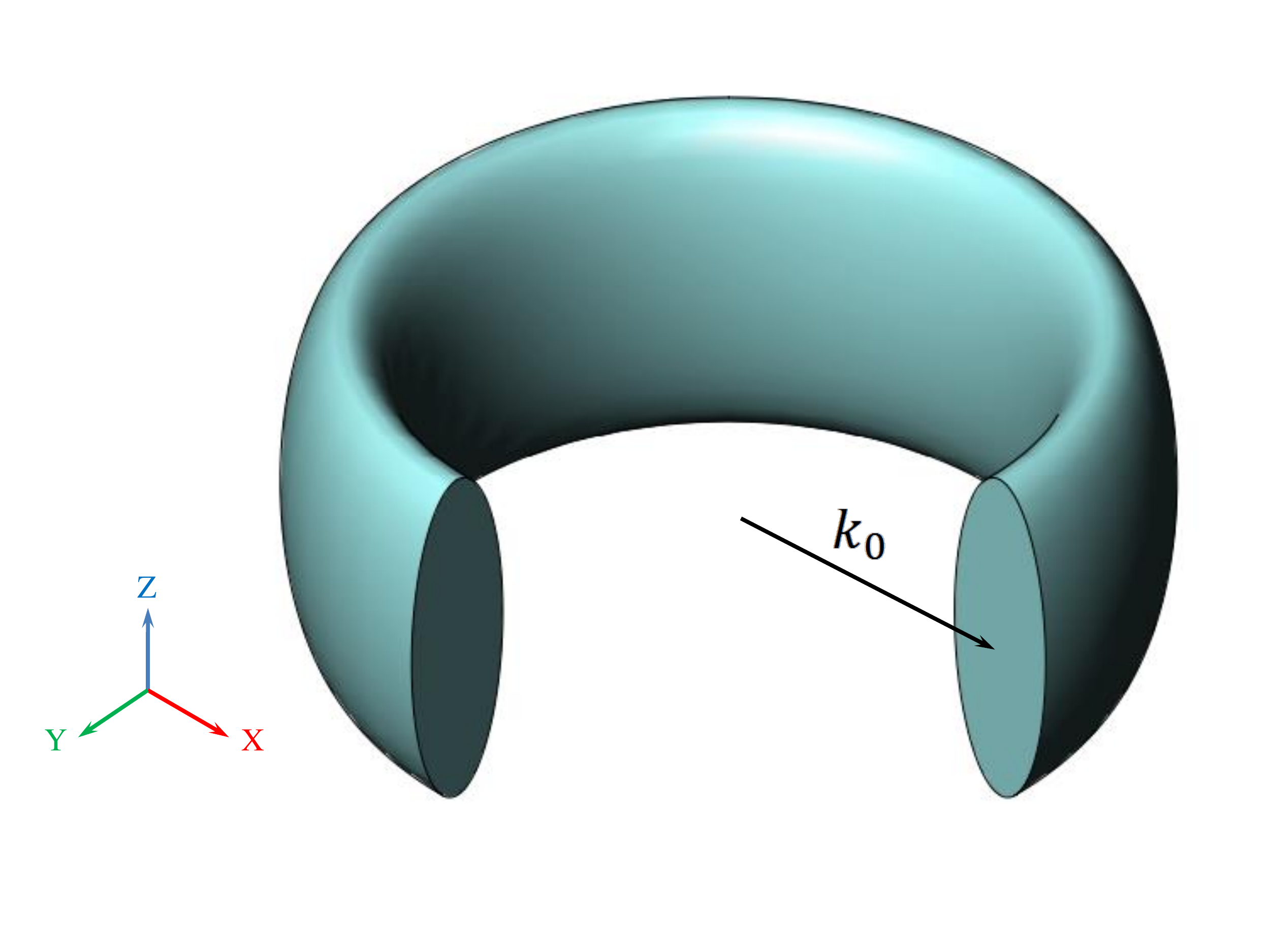} 
   \caption{An schematic scratch of the toroidal Fermi surface of a three dimensional nodal line semimetal. 
   \label{fig:fermi}}
\end{figure}

The low energy states of three dimensional nodal line semimetals could be modeled by different classes of effective continuum model Hamiltonians~\cite{fang_chinphys2016}. Here, we adapt a simple model~\cite{chan_prb2016,huh_prb2016,yan_prb2016}
\be\label{eq:hamil}
{\cal H}=\frac{\hbar^2}{2 m}\left(k_{\rho}^2-k_{0}^2\right)\hat{\tau}_{x}+\hbar v_z k_{z}\hat{\tau}_{y}~,
\ee
which describes the low energy states of a three dimensional nodal line semimetal in the absence of spin-orbit coupling, with a nodal ring at $k_z=0$-plane, with radius $k_0$ (see, Fig.~\ref{fig:fermi}). Here $m$ is the band mass in the $xy$-plane, $k_{\rho}=\sqrt{k_x^2+k_y^2}$ is the radial component of the wave vector, $v_z$ is a the Fermi velocity along the z direction, and $\hat{\tau}_x$ and $\hat{\tau}_y$ are respectively the $x$ and $y$ components of the Pauli matrices, acting on the pseudo-spin (\textit{i.e.}, orbital) degree of freedom.
The model Hamltonian~(\ref{eq:hamil}) provides a satisfactory low energy description for 3D NLSMs with SU(2) spin-rotation symmetry~\cite{chan_prb2016} and, as will be shown in the next section, is still simple enough to allow a fully analytical investigation of the optical response. 

\begin{figure}
\centering
\includegraphics[width=0.9\linewidth]{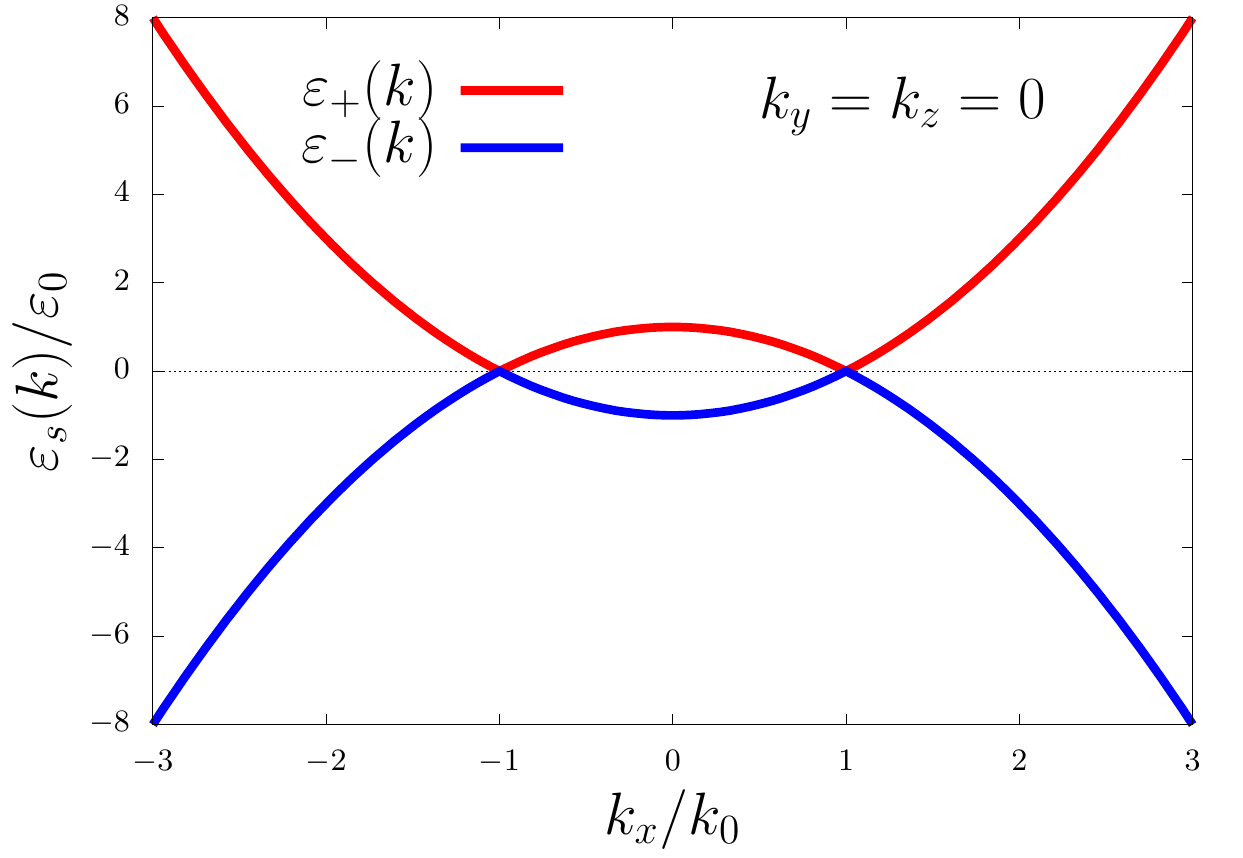}\\
\includegraphics[width=0.9\linewidth]{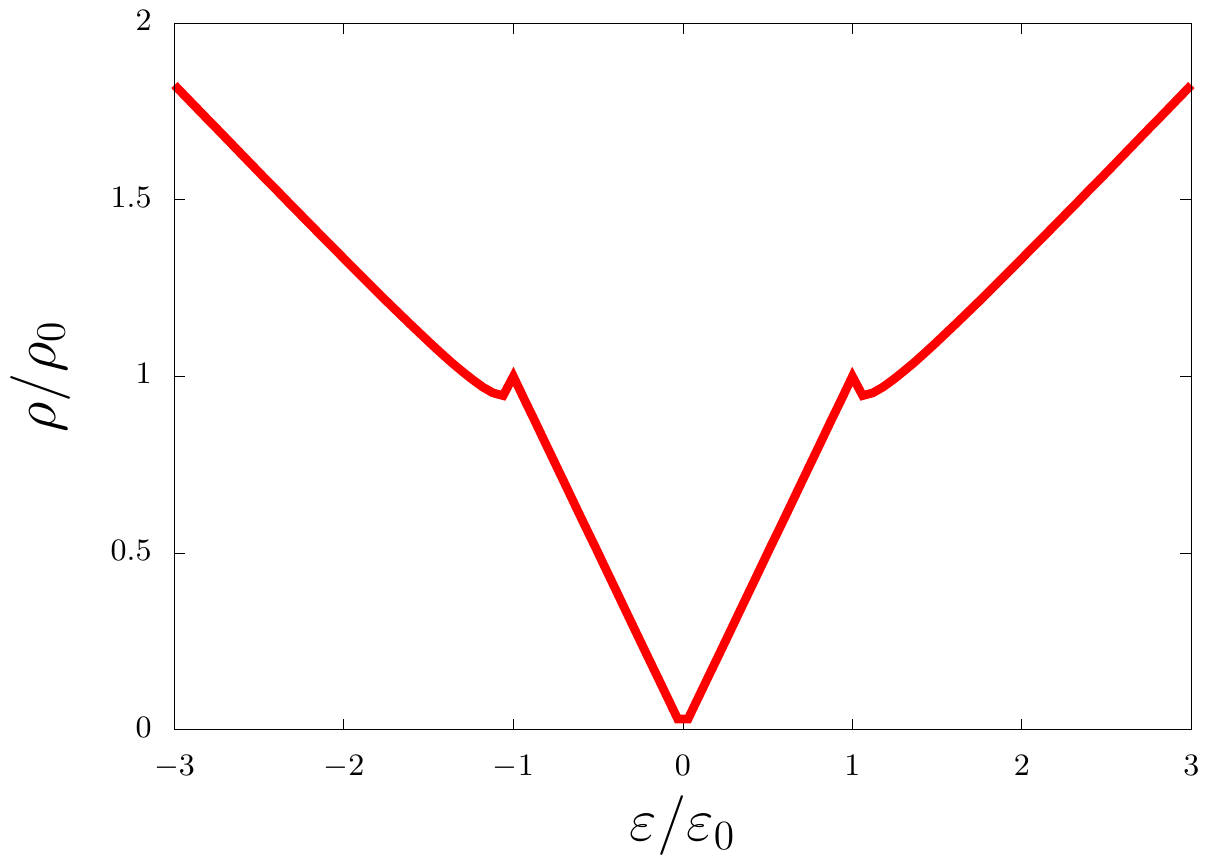}
\caption{
(Top) The low energy dispersions of a three dimensional nodal line semimetal (in units of $\varepsilon_0$) versus $k_x/k_0$, at $k_y=k_z=0$. 
(Bottom) The density of states of a three dimensional nodal line semimetal (in units of $\rho_0$) versus $\varepsilon/\varepsilon_0$.
\label{fig:eigen}}
\end{figure}

For notational convenience, we adopt the dimensionless units, where all the wave vectors are expressed in the units of $k_0$, and all the energies in the units of $\varepsilon_0=\hbar^2 k_0^2/( 2m)$. 
The eigenvalues of Hamiltonian~(\ref{eq:hamil}) are
\be\label{eq:eigen}
{\tilde \varepsilon}_{\kv,s} =s \sqrt{\left({\tilde k}_{\rho}^2-1\right)^2+ \gamma^2 {\tilde k}_{z}^2}~,
\ee
where ${\tilde \varepsilon}_{\kv,s}= \varepsilon_{\kv,s}/ \varepsilon_0$, ${\tilde k}_\rho=k_\rho/k_0$, ${\tilde k}_z=k_z/k_0$, and
$\gamma=2 m v_z/(\hbar k_0)$ is the dimensionless axial Fermi velocity, $s= + (-)$ labels the conduction (valance) band, and their corresponding eigenstates are given by
\be\label{eq:eigenstate}
\psi_{\kv,s}(\rv)
=\langle \rv | \kv,s\rangle
=\dfrac{1}{\sqrt{2V}}
\left(
\begin{array}{c}
1\\
s e^{i\theta_\kv }\\
\end{array} \right) e^{i\kv\cdot\rv}~,
\ee
where $V$ is the sample volume and
\be
\theta_\kv=\arctan\left(\frac{\gamma {\tilde k}_z}{{\tilde k}^2_\rho-1}\right)~.
\ee
The energy dispersion has been illustrated in Fig.~\ref{fig:eigen} (top panel). 
The density-of-states per unit volume (DOS)  of this system could be readily obtained as
\be\label{eq:dos}
\begin{split}
\rho(\varepsilon)&=\frac{g_s}{V}\sum_{\kv,s}\delta(\varepsilon-\varepsilon_{\kv,s})\\
&=\rho_0|{\tilde \varepsilon}|
\left[
1-\frac{1}{\pi}\Theta\left(|{\tilde \varepsilon}|- 1\right){\rm arcsec}\left(|\tilde  \varepsilon|\right)
\right]~,
\end{split}
\ee
where $g_s=2$ accounts for the real spin degeneracy, $\rho_0=k_0^3/(2\pi \varepsilon_0 \gamma)$, ${\tilde \varepsilon}=\varepsilon/\varepsilon_0$, and $\Theta(x)$ is the Heaviside step function.
Note that, as $\rho(\varepsilon  \to \infty)  \approx \rho_0(1/\pi+{\tilde \varepsilon}/2)$, 
the DOS has linear dependence on energy at both low and high energies [see, Fig.~\ref{fig:eigen} (bottom) for its full energy dependence].
The Fermi energy $\varepsilon_{\rm F}$ of an intrinsic (\textit{i.e.}, undoped) NLSM would lie at the touching line between valance and conduction bands. At low electron or hole doping, as long as $|\varepsilon_{\rm F}| < \varepsilon_0$, the Fermi surface has an anisotropic toroidal shape (see, Fig.~\ref{fig:fermi}), while at higher carrier densities, the Fermi surface becomes a spheroid. 

\section{Kubo formula for optical conductivity}\label{sec:kubo}
We calculate the longitudinal dynamical conductivity of a 3D NLSM within the linear response theory, using the Kubo formula for ac conductivity in the optical (\emph{i.e.}, long wavelength) limit~\cite{giuliani_vignale}
\be\label{eq:kubo}
 \sigma_{\alpha \beta}(\omega)
 =\frac{\hbar g_s}{iV}\sum_{\kv,s,s'}
\frac{f(\varepsilon_{\kv,s})-f(\varepsilon_{\kv,s'})}{\varepsilon_{\kv,s}-\varepsilon_{\kv,s'}}
  \frac{ j^{s,s'}_{\kv,\alpha}j^{s',s}_{\kv,\beta}}{\hbar\omega +\varepsilon_{\kv,s}-\varepsilon_{\kv,s'}+i\eta}
 ~,
\ee
where $\alpha$ and $\beta$ refer to three spacial directions x, y, and z, $j^{s,s'}_{\kv,\alpha}=\langle \kv,s | j_\alpha|\kv,s' \rangle$ are the matrix elements of the paramagnetic component of the current-density operator ${\bf j}=-i e[H , \rv]/\hbar$, along the $\alpha$-direction, $f(\varepsilon)=1/[e^{\beta(\varepsilon-\mu)}+1]$ is the Fermi-Dirac distribution function at the inverse temperature $\beta=1/(k_{\rm B}T)$ and the chemical potential $\mu$, and $\eta=\hbar/\tau$ accounts for the lifetime of energy states. 
This lifetime depends on the relevant scattering mechanisms, but we will treat it as a constant phenomenological parameter, and take the $\eta \to 0^+$ limit for clean systems.
The cartesian components of the paramagnetic current operator of a 3D nodal line semimetal, described by Hamiltonian~(\ref{eq:hamil}), read
\be
\begin{split}
j_x &=- \frac{e\hbar}{m}  k_x \hat{\tau}_x~, \\
j_y &=- \frac{e \hbar }{m} k_y\hat{\tau}_x~, \\
j_z &=- e v_z \hat{\tau}_y~,
\end{split}
\ee
and the matrix elements of these operators in the basis~(\ref{eq:eigenstate}) of Hamiltonian are 
\be
 \begin{split}
j^{s,s'}_{\kv,x} & = \langle \kv,s|j_x|\kv,s'\rangle=- \frac{e \hbar}{2 m}k_x\left(s e^{-i \theta_{\kv}}+s' e^{i \theta_{\kv}}\right)~,\\
j^{s,s'}_{\kv,y} & = \langle \kv,s|j_y|\kv,s'\rangle=- \frac{e \hbar}{2 m} k_y \left(s e^{-i \theta_{\kv}}+s' e^{i \theta_{\kv}}\right)~,\\
j^{s,s'}_{\kv,z}&= \langle \kv,s|j_z|\kv,s'\rangle=- \frac{i e v_z}{2}\left(s e^{-i \theta_{\kv}}- s'  e^{i \theta_{\kv}}\right)~.
 \end{split}
 \ee
It could be easily shown that only two independent components of the $3\times 3$ conductivity matrix (\ref{eq:kubo}) are nonzero: 
$\sigma_{zz}(\omega)$ and $\sigma_{xx}(\omega)=\sigma_{yy}(\omega)$. 
For a clean system (\textit{i.e.}, in the $\eta \to 0$ limit), the real part of the optical conductivity reads
\be\label{eq:re_kubo}
\begin{split}
\Re e\, \sigma_{\alpha \alpha}(\omega,\eta=0)
 =-\frac{\pi g_s}{V\omega}&\sum_{\kv,s,s'}
 \left|j^{s,s'}_{\kv,\alpha}\right|^2
 \delta \left(\hbar\omega +\varepsilon_{\kv,s}-\varepsilon_{\kv,s'} \right)\\
  \times & \left[f(\varepsilon_{\kv,s}+\hbar \omega)-f(\varepsilon_{\kv,s})\right]
 ~.
 \end{split}
\ee
For a doped system (\textit{i.e.}, at finite $\eta$), the real part of the optical conductivity could be obtained from the clean system conductivity (\ref{eq:re_kubo}), through
\be\label{eq:sigma_eta}
\Re e\, \sigma_{\alpha \alpha}(\omega,\eta)=\frac{\hbar \eta}{\pi}\int\mathrm{d}\omega'\frac{\Re e\, \sigma_{\alpha \alpha}(\omega',\eta=0)}{(\hbar \omega-\hbar \omega')^2+\eta^2}~.
\ee
In the following, we derive both intraband ($s'=s$) and interband ($s'=-s$) contributions to the optical conductivity of a NLSM along the $x$ and $z$ directions. 
For notational brevity, explicit references to $\eta$ in the argument of optical conductivity will be mainly dropped. 
Whether each result is for a clean or an impure system, would be clear from their expressions.

\subsection{Drude conductivity}
The intraband contribution to the optical conductivity of a clean system arises from the $s'=s$ terms of Eq.~(\ref{eq:re_kubo}), which could be written as
\be\label{eq:sigma_d}
\Re e\, \sigma^{\rm D}_{\alpha \alpha}(\omega)
 =D_\alpha \delta(\omega)~,
\ee
and for a doped system, from Eq.~(\ref{eq:sigma_eta}) we find the familiar lorentzian expression
\be\label{eq:sigma_d_doped}
\Re e\, \sigma^{\rm D}_{\alpha \alpha}(\omega,\eta)=\frac{D}{\pi}\frac{\hbar \eta}{\hbar^2\omega^2+\eta^2}~.
\ee
Here
\be\label{eq:drude}
D_\alpha
 =-\frac{\pi g_s }{V}\sum_{\kv,s}
 |j^{s,s}_{\kv,\alpha}|^2
 \left.\frac{\partial f(\varepsilon)}{\partial \varepsilon}\right|_{\varepsilon=\varepsilon_{\kv,s}} ~,
\ee
is the Drude weight or charge stiffness in the $\alpha$ direction.
For the x-component of the current operator in the cylindrical coordinates, we have
\be
|j^{s,s}_{\kv,x}|^2=\frac{e^2\hbar^2}{m^2} k^2_\rho \cos^2 \phi_\kv \cos^2 \theta_\kv~,
\ee 
and then for the radial component of the Drude weight, using the fact that $-\partial f(\varepsilon)/\partial \varepsilon =\delta(\mu-\varepsilon)$ at zero temperature, and considering $\mu>0$, we obtain
\be\label{eq:drudx}
D_x=D_x^0{\tilde \mu}
\left\{1+\frac{\Theta({\tilde \mu}-1)}{\pi} 
\left[\frac{4 {\tilde \mu}^2-1}{3 {\tilde \mu}^2} \sqrt{{\tilde \mu}^2-1}-{\rm arcsec}({\tilde \mu})
\right] \right\}~,
\ee
where ${\tilde \mu}=\mu/\varepsilon_0$ and $D_{x}^{0}=e^2 k_0^3/(4 m\gamma)$.
Note that at low carrier doping (\emph{i.e.}, $ \mu < \varepsilon_0$), this radial Drude weight linearly depends on the chemical potential. On the other hand, when the chemical potential is very high, or the nodal ring radius is very small (\emph{i.e.}, $\varepsilon_0 \ll \mu$), the Drude weight becomes independent of $k_0$, and quadratically increases with $\mu$.   

\begin{figure}
\includegraphics[width=0.9\linewidth]{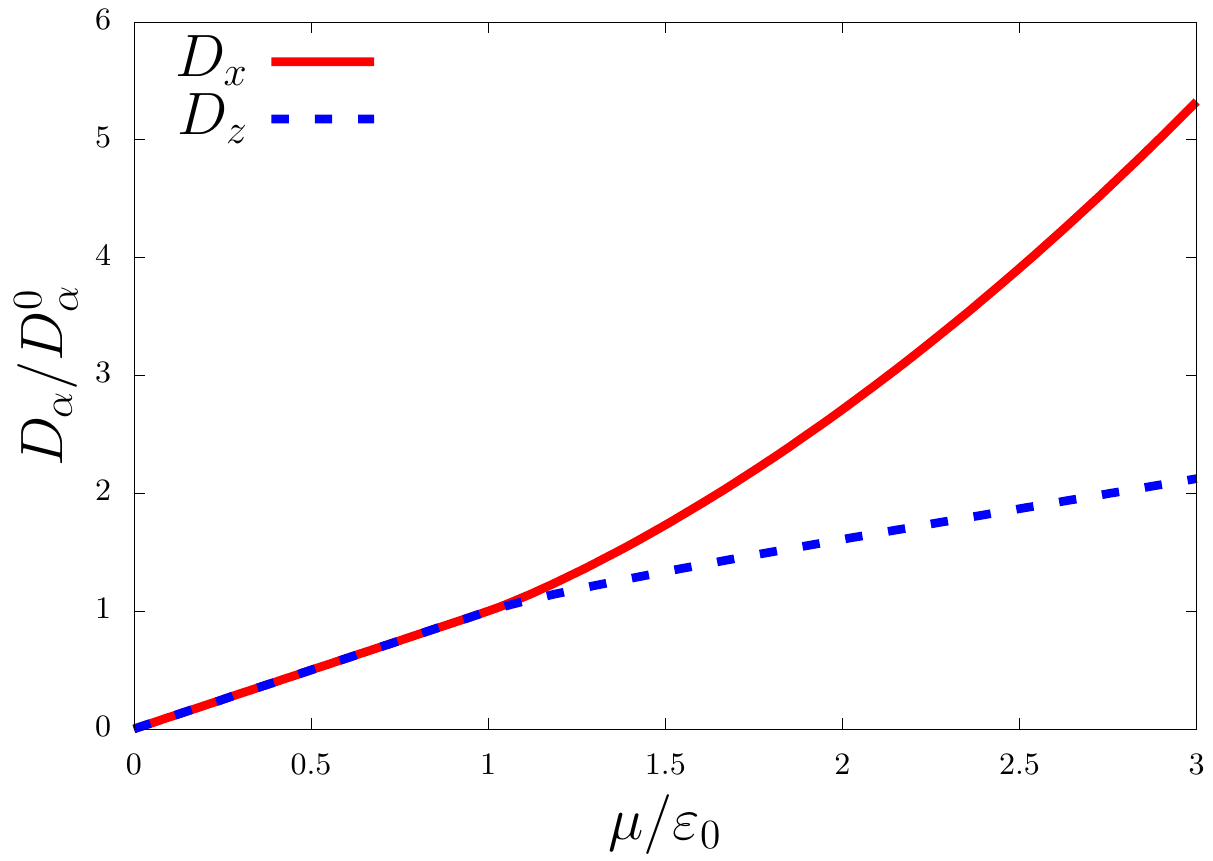}
\caption{The zero temperature Drude weight of a three dimensional nodal line semimetal, along the radial ($x$) and axial ($z$) directions, versus the chemical potential $\mu/\varepsilon_0$. Recall that $D^0_z=D^0_x\gamma^2/2$.
\label{fig:drude}}
\end{figure}

Along the z-direction, the matrix elements of current operator read
\be
|j^{s,s}_{\kv,z}|^2=e^2v_z^2 \sin^2 \theta_\kv~,
\ee
and for the axial component of the Drude weight at zero temperature, we find
\be\label{eq:drude_z}
D_z=D_z^0{\tilde \mu}\left\{
1+\frac{\Theta({\tilde \mu}-1)}{\pi}
\left[\frac{\sqrt{ {\tilde \mu}^2-1}}{{\tilde \mu}^2} -{\rm arcsec}({\tilde \mu})\right]\right\}
~,
\ee
where  $D_z^0=\gamma^2 D_x^0/2$.
The Drude weight along the z-direction has a linear dependence on chemical potential at both low and high dopings. At low doping (\emph{i.e.}, $ \mu < \varepsilon_0$), $D_z$ is independent of the nodal ring radius $k_0$. When the chemical potential is very large, or the nodal ring radius is very small (\emph{i.e.}, $\varepsilon_0 \ll \mu$), the Drude weight approaches $D^0_z({\tilde \mu}/2+2/\pi)$. 
In Fig.~\ref{fig:drude}, we have illustrated the full chemical potential dependence of the Drude weights along the radial and axial directions.  We should note that the Drude weights of clean Weyl and Dirac semimetals have quadratic dependence on chemical potential~\cite{tabert_prb2016_1}.  
At finite temperatures, Drude weights could be obtained from the numerical solution of
\be
D_\alpha(\mu,T)=-\int \mathrm{d} \varepsilon \frac{\partial f(\varepsilon)}{\partial \varepsilon} D_\alpha(\varepsilon,T=0)~.
\ee
\subsection{Interband conductivity}
The interband contribution to the optical conductivity of a clean system arises from the $s'=-s$ terms of Eq.~(\ref{eq:re_kubo})
\be\label{eq:re_intra}
\Re e\, \sigma^{{\rm int}}_{\alpha \alpha}(\omega)=
\frac{\pi  g_sG(\omega)}{V\omega}
\sum_{\kv,s}
\left|j^{s,-s}_{\kv,\alpha}\right|^2
\delta \left(\hbar\omega -2\varepsilon_{\kv,s}\right)~.
\ee
Here we have considered $\omega>0$, and all the temperature dependence is carried into the thermal factor
\be
\begin{split}
G(\omega)&=f(\hbar\omega/2)-f(-\hbar\omega/2)\\
&=\frac{\sinh(\beta \hbar \omega/2)}{\cosh(\beta \hbar \omega/2)+\cosh(\beta \mu)}~.
\end{split}
\ee
The interband matrix elements of the current operators are
\be
|j^{+,-}_{\kv,x}|^2=\frac{e^2\hbar^2}{m^2} k^2_\rho \cos^2 \phi_\kv \sin^2 \theta_\kv~,
\ee 
and
\be
|j^{+,-}_{\kv,z}|^2=e^2v_z^2 \cos^2 \theta_\kv~,
\ee
and the corresponding components of the interband conductivities, after some straightforward algebra could be obtained as  
\begin{widetext}
\be
\Re e\, \sigma^{\rm int}_{xx}(\omega)=\sigma^0_{xx} G(\omega)
\left\{1-\frac{\Theta({\tilde \omega} -1)}{\pi}\left[{\rm arcsec}\left({\tilde \omega}\right)-\frac{2 \tilde \omega^2+1}{3 \tilde \omega^2} \sqrt{{\tilde \omega}^2-1}\right]\right\}~,
\ee
and
\be
\Re e\, \sigma^{\rm int}_{zz}(\omega)=\sigma^0_{zz} G(\omega)
\left\{1-\frac{\Theta({\tilde \omega} -1)}{\pi}\left[{\rm arcsec}\left({\tilde \omega}\right)+\frac{\sqrt{{\tilde \omega}^2-1}}{{\tilde \omega}^2}\right]\right\}~,
\ee
\end{widetext}
where ${\tilde \omega}=\hbar \omega/(2\varepsilon_0)$ and $\sigma^0_{\alpha\alpha}=\hbar D^0_{\alpha}/(4\varepsilon_0)$.
Note that at zero temperature $G(\omega)=\Theta(\hbar\omega-2\mu)$, and the intraband conductivities naturally become zero for $0<\hbar \omega < 2 \mu$, due to the Pauli blocking. For low carrier concentrations, they also have plateaus for frequencies $2\mu<\hbar \omega<2 \varepsilon_0$, where $\Re e\, \sigma^{\rm int}_{\alpha\alpha}(\omega)=\sigma^0_{\alpha\alpha}$. At high frequencies, the radial conductivity $\Re e\, \sigma^{\rm int}_{xx}(\omega)$ linearly increases with frequency, while the axial component $\Re e\, \sigma^{\rm int}_{zz}(\omega)$ approaches the constant value $\sigma^0_{zz}/2$.
\begin{figure}
\includegraphics[width=0.9\linewidth]{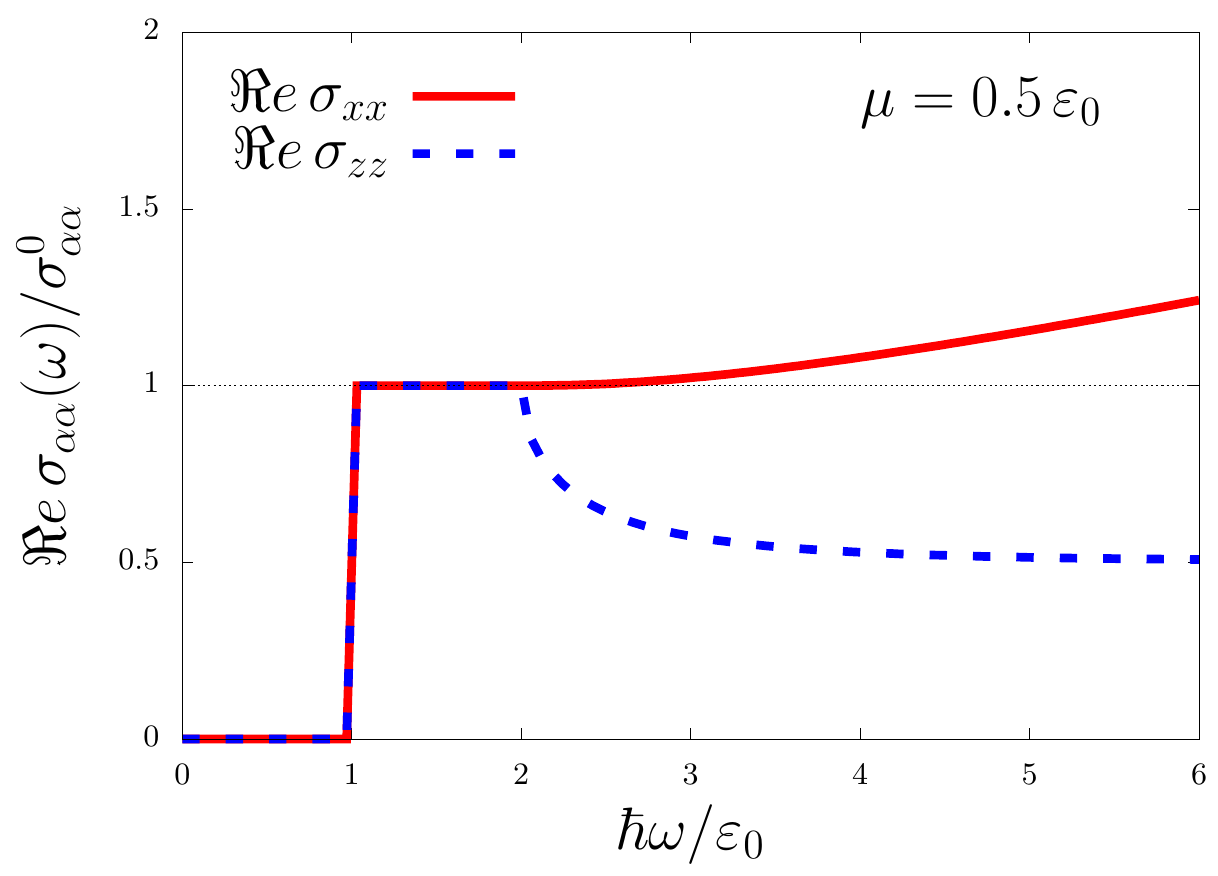}
\caption{The frequency dependence of the interband optical conductivity of a clean three dimensional nodal line semimetal, along the radial ($\sigma_{xx}$) and axial ($\sigma_{zz}$) directions at zero temperature and for $\mu=0.5\, \varepsilon_0$. \label{fig:sigmaxz}}
\end{figure}
In Fig.~\ref{fig:sigmaxz}, we compare the frequency dependence of $x$ and $z$ components of the intraband conductivities of a clean 3D NLSM at zero temperature, and for $\mu=0.5\, \varepsilon_0$.

\section{Optical conductivity of 2D nodal line semimetals}\label{sec:2D-NLSM}
Similar to 3D NLSMs, also in 2D nodal line semimetals the conduction and valance bands touch along a nodal ring. We model such a system with a very simple effective Hamiltonian~\cite{li_prb2016}
\be\label{eq:hamil2D}
{\cal H}_{\rm 2D}=\frac{\hbar^2}{2 m}\left(k_{\rho}^2-k_{0}^2\right)\hat{\tau}_{x}~,
\ee
which could be obtained from the $v_z\to 0$ limit of Hamiltonian~(\ref{eq:hamil}).
Its energy spectrum and eigenstates, respectively are 
\be\label{eq:eigen2D}
{\tilde \varepsilon}_{\kv,s} =s |{\tilde k}_{\rho}^2-1|~,
\ee 
and 
\be\label{eq:eigenstate2D}
\psi_{\kv,s}(\rv)
=\langle \rv | \kv,s\rangle
=\dfrac{1}{\sqrt{2A}}
\left(
\begin{array}{c}
1\\
s \\
\end{array} \right) e^{i\kv\cdot\rv}~,
\ee
where $A$ is the  surface area. The DOS of a 2D nodal line semimetal could be readily obtained as
\be\label{eq:dos2D}
\rho_{\textrm{2D}}(\varepsilon)=\rho^0_{\textrm{2D}}
\left[2-\Theta(|{\tilde \varepsilon}|-1)\right]~,
\ee
where $\rho^0_{\textrm{2D}}=m/(\pi\hbar^2)$, is identical to the DOS of a single band  conventional 2D electron gas~\cite{giuliani_vignale}. 
The matrix elements of current operator $j_x =-  e\hbar  k_x \hat{\tau}_x/m$, in the basis of Hamiltonian~(\ref{eq:hamil2D}) are
\be
j^{s,s'}_{\kv,\alpha}  = \langle \kv,s |j_{\alpha}|\kv,s'\rangle=- \frac{e \hbar}{2 m}k_{\alpha}\left(s+s' \right)~.
 \ee
As $j^{+,-}_{\kv,\alpha}=0$, the interband contribution to the optical conductivity is identically zero, and the Drude weight reads
\be\label{eq:drude2D}
D_x=D^0_{\textrm{2D}} \left[ 2+ \Theta({\tilde \mu}-1) \left({\tilde \mu}-1\right) \right]~,
\ee
with $D^0_{\textrm{2D}}=e^2 k_0^2/(2 m)$. It is interesting to note that, the Drude weight is independent of the carrier density at low chemical potentials. The conductivities of clean and impure systems would still be given by Eqs.~(\ref{eq:sigma_d}) and (\ref{eq:sigma_d_doped}), respectively, after replacing the 2D Drude weight from Eq.~(\ref{eq:drude2D}).

\section{Summary and Discussion}\label{sec:summary}
In summary, we have investigated the optical conductivity of two and three-dimensional nodal line semimetals using the Kubo formula.
In two-dimensional NLSMs, for the particular model, we have considered here, the interband conductivity vanishes and the Drude weight is independent of the electron density for low chemical potentials. The density dependence of the dc conductivity in this regime will depend on the microscopic origin of the impurity scattering and could be investigated using \textit{e.g.}, the relaxation time approximation and its generalization for anisotropic systems~\cite{amir_jpcm2015}.
The anisotropic shape of the Fermi surface of three-dimensional NLSMs results in their anisotropic Drude weights and optical conductivities. 
The anisotropic Drude weight would be responsible for the anisotropic dispersion of collective modes in these materials~\cite{yan_prb2016}. Effects of many-body correlations on the Drude weight and collective modes~\cite{saeed_prb2011} of nodal point and nodal line semimetals would be a very interesting open problem.

A model Hamiltonian, very similar to our Eq.~(\ref{eq:hamil}), has been fitted to the density functional theory (DFT) results for the band structure of Ca$_3$P$_2$ by Chan \textit{et al.},~\cite{chan_prb2016}. 
In terms of our model parameters, that parametrization gives $k_0\approx 0.206\, \AA^{-1}$, $\varepsilon_0\approx 0.184 \, eV$, and $\gamma \approx 2.80$. Therefore, for the anisotropy of Drude weight and interband conductivity of Ca$_3$P$_2$, our findings predict $D^0_z \approx 3.92 D^0_x$ and $\sigma^0_{zz} \approx 3.92 \sigma^0_{xx}$, respectively.

Finally, we should note that throughout this paper we have used a two-band low energy model Hamiltonian. Therefore, our results for the Drude weight and optical conductivity would be reliable only at low chemical potentials and low frequencies. 
At high frequencies or large chemical potentials, the higher order terms in the wave vector and the effects of other bands may not be negligible. The contribution of these terms is expected to substantially affect the optical response at high frequency and large chemical potential. In particular, the linear dependence of the radial interband conductivity at high frequency could be an artifact of our simple low energy model.

\appendix*
\section{Carrier density and the Fermi energy}\label{sec:fermi}
At zero temperature, the chemical potential is equal to the Fermi energy, which is the highest occupied energy state and is related to the density of doped carriers $n_e$. Considering only the electron doped case (\textit{i.e}, $\varepsilon_{\rm F}>0$), we can write
\be
n_e=\int_0^{\varepsilon_{\rm F}} \mathrm{d}\varepsilon\, \rho(\varepsilon)~.
\ee
In two dimensional nodal line semimetals, the density of states is constant [see, Eq.~(\ref{eq:dos2D})], and we find
\be
\varepsilon_{\rm F}=\frac{\pi \hbar^2}{2 m}\left[ n_e+ \Theta(n_e-n^{0}_{\textrm{2D}})\left(n_e-n^{0}_{\textrm{2D}}\right)\right]~,
\ee
where $n^{0}_{\textrm{2D}}=k_0^2/\pi$, corresponds to the density where $\varepsilon_{\rm F}=\varepsilon_0$.
In 3D nodal line semimetals, the complicated energy dependence of DOS, makes it difficult to find an analytic expression for Fermi energy, over the whole range of electron concentrations. But, for low and very high dopings, one can obtain
\be
\left\{
\begin{array}{c}
\varepsilon_{\rm F}=\sqrt{2\varepsilon_0 n_e/\rho_0}~~~~~ \textrm{for}~~~~n_e <n^{0}_{\textrm{3D}}~,\\ \\
\varepsilon_{\rm F}=\sqrt{4\varepsilon_0 n_e/\rho_0}~~~~~ \textrm{for}~~~~n_e \gg n^{0}_{\textrm{3D}}~,
\end{array}\right. ,
\ee
where $n^{0}_{\textrm{3D}}=k_0^3/(4 \pi \gamma)$ is the density in which $\varepsilon_{\rm F}=\varepsilon_0$.

\end{document}